\documentclass[prd, 10pt,aps,nofootinbib, floatfix, notitlepage,twocolumn,longbibliography]{revtex4-1}
              
\pdfoutput=1

\usepackage{amsmath,amsfonts,amssymb}
\usepackage{mathrsfs}
\usepackage{graphicx}
\usepackage[english]{babel} 
\usepackage{url} 
\usepackage{color}
\usepackage{bbold}
\usepackage{adjustbox}
 \usepackage[utf8]{inputenc} 
\usepackage[colorlinks=true,citecolor=blue,urlcolor=blue]{hyperref}

\newcommand{\M}{\Lambda}
\newcommand{\mpl}{M_{\mbox{\tiny{Pl}}}}

\begin{document}

\title{Ultralight Scalar Decay and the Hubble Tension}

\author{Mark Gonzalez}
\email{mark.gonzalez@tufts.edu}
\author{Mark P.~Hertzberg}
\email{mark.hertzberg@tufts.edu}
\author{Fabrizio Rompineve}
\email{fabrizio.rompineve@tufts.edu}
\affiliation{Institute of Cosmology, Department of Physics and Astronomy, Tufts University, Medford, MA 02155, USA
\looseness=-1}

\date{\today}

\begin{abstract}
We examine whether the Hubble tension, the mismatch between early and late measurements of $H_0$, can be alleviated by ultralight scalar fields in the early universe, and we assess its plausibility within UV physics. Since their energy density needs to rapidly redshift away, we explore decays to massless fields around the era of matter-radiation equality. We highlight a concrete implementation of ultralight pseudo-scalars, axions, that decay to an abelian dark sector. This scenario circumvents major problems of other popular realizations of early universe scalar models in that it uses a regular scalar potential that is quadratic around the minimum, instead of the extreme fine-tuning of many existing models. The idea is that the scalar is initially frozen in its potential until $H\sim m$, then efficient energy transfer from the scalar to the massless field can occur shortly after the beginning of oscillations due to resonance. We introduce an effective fluid model which captures the transition from the frozen scalar phase to the radiation dark sector phase. We perform a fit to a combined Planck 2018, BAO, SH$_0$ES and Pantheon supernovae dataset and find that the model gives $H_0=69.9_{-0.86}^{+0.84}$ km/s/Mpc with $\Delta\chi^2 \approx -9$ compared to $\Lambda$CDM; while inclusions of other data sets may worsen the fit. Importantly, we find that large values of the coupling between fields is required for sufficiently rapid decay: For axion-gauge field models $\phi F\tilde{F}/\Lambda$ it requires $\Lambda\lesssim f/80$, where $2\pi f$ is the field range. We find related conclusions for scalar-scalar models $\sim\phi\,\chi^2$ and for models that utilize perturbative decays. We conclude that these sorts of ultralight scalar models that purport to alleviate the Hubble tension, while being reasonable effective field theories, require features that are difficult to embed within UV physics.
\end{abstract}

\maketitle

\section{Introduction}

The standard cosmological model is relatively simple (albeit possibly hiding deep conceptual questions): the Universe contains (visible and dark) matter, radiation, and dark energy (possibly a cosmological constant). Observations of the Cosmic Microwave Background (CMB) suggest the addition of an early inflationary phase, likely driven by a scalar field, which ends with decays into matter and radiation. The physics of this so-called reheating process can be complex; for instance, in many scenarios the visible Universe is produced during the inflaton oscillations after inflation by means of a very violent resonant decay process, also referred to as preheating. This complexity is nonetheless most likely inaccessible observationally, since inflation supposedly occurred at very high energies/redshifts.

At much lower energies, the simplicity of the $\Lambda$CDM scenario has instead been called into question by some observations of the late Universe. In particular, measurements of the Hubble expansion rate by means of the distance ladder method exhibit tension with the extraction of this parameter from CMB data. Among late-time measurements, the most precise has been performed by the SH$_{0}$ES team using Supernovae data and finds $H_{0}=74.03\pm 1.42~\text{km/s/Mpc}$~\cite{Riess:2019cxk}. In contrast, the Planck collaboration reports $H_{0}=67.27\pm 0.60$ km/s/Mpc~\cite{Aghanim:2018eyx} from the fit of $\Lambda$CDM to CMB power spectra. The two approaches disagree at the 4.4$\sigma$ level. Other independent local measurements mostly tend to agree with the SH$_{0}$ES measurement rather than with the value inferred by Planck (see however~\cite{Freedman:2019jwv} for a measurement which agrees with both late and early time measurements and e.g.~\cite{Verde:2019ivm} for a complete review), and similarly the combination of CMB and Baryon Acoustic Oscillations (BAO) data gives a result in very good agreement with the aforementioned Planck value~\cite{Aghanim:2018eyx}. 

Since no explanation of the tension based on systematic errors has arisen at the time of writing, it is legitimate to think about physics beyond $\Lambda$CDM which could help in reconciling the measurements by changing the results of the fits to CMB data. Several ingredients have been proposed and among them the two most popular additions are as follows. (a) dark radiation, which we will refer to as the $\Delta N_{\text{eff}}$ model (see~\cite{Aghanim:2018eyx},~\cite{DEramo:2018vss} for a realization of this scenario involving the QCD axion and~\cite{Vagnozzi:2019ezj, Barker:2020gcp} for further discussions and scenarios of dark radiation in relation to the Hubble tension). This model has the advantage of only one extra parameter and is reasonably well motivated from the UV point of view. However, it only leads to a very mild improvement in $\Delta\chi^2$. Given that the improvement in $\Delta\chi^2$ from $\Delta N_{\text{eff}}$ is rather small, there has been an interest in more exotic new physics, especially (b) the so-called early dark energy (EDE) models. In this case, an ultralight scalar field is introduced that initially has equation of state $w\approx-1$, but then is assumed to evolve in a strange potential that (despite its name) leads to behavior that is essentially the opposite of that of dark energy at later times: it redshifts away very quickly around matter-radiation equality due to an unusual potential that has no quadratic term around its minimum and hence no mass \cite{Poulin:2018cxd, Agrawal:2019lmo, Lin:2019qug, Smith:2019ihp} (see also~\cite{Braglia:2020bym} for a supergravity inspired model). These models have been argued to have a better $\Delta\chi^2$. However, from the point of view of fundamental physics they are peculiar, since the vanishing of the mass around the minimum, without explanation, requires fine-tuning.

Given this, our interest in this work is to investigate whether there exist models that can (i) alleviate the Hubble tension with a good $\Delta\chi^2$ and also (ii) be plausible from the point of view of fundamental physics. So we develop and investigate here a class of models that 
features some of the advantages of both the $\Delta N_{\text{eff}}$ and EDE models, but may improve on their obvious downsides. 
We consider the possibility that the Universe contains a dark sector, with an originally slowly-rolling scalar field which then rapidly decays into massless (or essentially massless) particles. The most basic version of this is to assume the scalar is an axion that resonantly decays to some dark radiation (see \cite{Agrawal:2017eqm, Kitajima:2017peg} for studies of this decay to reduce the relic abundance of the QCD axion, \cite{Hertzberg:2018zte,Hertzberg:2020dbk} for applications to axion dark matter clumps, \cite{Berghaus:2019cls, Niedermann:2019olb,  Davoudiasl:2020opf, Niedermann:2020dwg} for EDE models which similarly to ours feature a scalar field which decays to radiation, and other decaying models include \cite{Berezhiani:2015yta,Gu:2020ozv}). Our primary interest is to determine the parameter space required for this class of constructions, and then to comment on its status within fundamental physics. 

To set ideas, we consider as a minimal low-energy content of this sector an ultralight scalar with mass around $m\gtrsim H_{\text{eq}}\sim 10^{-28}~\text{eV}$ and $f\lesssim 10^{17}~\text{GeV}$, such that its initial energy density is $\lesssim 10\%$ of the radiation energy density around matter-radiation equality, and an abelian gauge field (for a qualitative discussion of resonant decay to scalar fields rather than gauge fields in the context of the Hubble tension see~\cite{Kaloper:2019lpl}). This setup is analogous in many respects to that of preheating after inflation, the crucial difference being obviously in the relevant energy scale (eV rather than $10^{15}$ GeV!). Because of this, such a \emph{decaying ultralight scalar} (dULS) scenario would be interesting in that, if viable, it could observationally probe the physics of resonant decay via CMB extraction of the Hubble parameter. 

This framework combines aspects of the EDE and $\Delta N_{\text{eff}}$ scenarios, in that we use an initially overdamped axion field as well as dark radiation, without suffering from the extreme fine-tuning of the EDE scenario. Indeed, we make use of a standard axion potential which is quadratic around its minimum. However, in order to achieve a sufficiently rapid energy transfer from the axion to the gauge field, a large coupling between the two species is required, which we will quantify. While this does not invalidate the effective field theory which we will be using, achieving this regime in a concrete UV setup is highly non-trivial. Furthermore, similarly to other EDE models, this scenario does not address the question of why a dynamical EDE transition should occur around the epoch of matter radiation equality (see instead e.g. \cite{Lin:2018nxe, Rossi:2019lgt, Sola:2019jek, Sakstein:2019fmf, Zumalacarregui:2020cjh, Ballesteros:2020sik, Braglia:2020iik} for a partial list of modified gravity and other scenarios which address this coincidence problem). We also find related requirements for large couplings in other field theories.

The aim of this work is to outline the above class of model in detail, to make a first step towards a realistic quantitative assessment of its ability to resolve the Hubble tension, and, importantly, to assess its plausibility within fundamental physics. Since a full numerical implementation of the two-field system in a cosmological Boltzmann code is challenging, we build an effective fluid model which mimics the evolution of the axion background and its decay to dark gauge fields. We then perform a fit of this simplified fluid model to a cosmological dataset which combines early and late-time measurements. While we are mainly motivated by axion resonant decay, our fluid parameterization may also be applied to a general class of decaying ultralight scalar models, such as those involving perturbative decays and/or different field content.

This paper is organized as follows: 
In Section \ref{sec:axion} we introduce the axion-dark-photon model.
In Section \ref{sec:resonant} we investigate the energy transfer from the axion to the dark-photons and the required parameters. 
In Section \ref{sec:other} we discuss other related models and their required parameters. 
In Section \ref{sec:fluid} we describe an effective single fluid model of the combined system. 
In Section \ref{sec:fit} we compare the model to cosmological datasets. 
Finally, in Section \ref{sec:conclusions} we offer our conclusions including comments on fundamental physics.

\section{Axion and Dark Radiation Models}
\label{sec:axion}

We focus on a simple model which describes a rapid energy transfer from an ultralight axion field to an Abelian gauge field $A_\mu$ which belongs to a possibly larger dark sector.
\begin{equation}
\label{eq:lagrangian}
\mathcal{L}=-\frac{1}{4}F_{\mu\nu}F^{\mu\nu}+{1\over2}\partial_{\mu}\phi\partial^\mu\phi-\frac{g_{\phi A}}{4}\phi F_{\mu\nu}\tilde{F}^{\mu\nu}-V(\phi),
\end{equation}
where $\tilde{F}^{\mu\nu}=\frac{1}{2}\epsilon^{\mu\nu\alpha\beta}F_{\alpha\beta}$ is the dual of the dark $U(1)$ field strength tensor. Note that the first three terms here respect the continuous shift symmetry of an axion $\phi\to\phi+\phi_0$. 
However, the final term is a potential $V(\phi)$ which breaks the shift symmetry; its specific choice is quite important. Firstly, it should respect the discrete periodicity of the axion $\phi\to\phi+2\pi f$, where $f$ is the so-called axion decay constant and sets the axion's field range. Such potentials that break a continuous shift symmetry can be generated by non-perturbative effects, including instantons. In general, such a potential can be expanded in harmonics as follows
\begin{equation}
\label{eq:potentialharmonic}
V(\phi)=m^2 f^2 \sum_n \,c_n\,\cos\left(n\phi\over f\right)
\end{equation}
where $c_n$ are the dimensionless amplitude of the $n^{th}$ harmonic, and we have extracted out a mass scale $m$ in front for convenience. 

In the literature on EDE, some work \cite{Poulin:2018cxd} has used the potential $V\propto (1-\cos(\phi/f))^3$ in order to provide a good fit to data. This requires the conspiracy among harmonics: $c_2/c_1=-2/5$ and $c_3/c_1=1/15$, with $c_4=c_5=\ldots =0$, which appears to be an extreme fine-tuning. By assuming such a conspiracy without justification, such a model deserves a big penalty in a fair $\Delta\chi^2$ analysis. 

Instead in this paper, we make the following much more standard assumption: the harmonics organize into the dilute instanton approximation in which there is a hierarchy: $c_1\gg c_2,\,c_3,\ldots$ and we keep the leading harmonic and ignore the others. Without loss of generality, we set $c_1=-1$, and we then add a constant, namely $V_0\approx m^2 f^2$ to ensure the late-time vacuum energy is small (the cosmological constant problem is clearly not addressed by ours or any other known models). This leads to the standard axion potential
\begin{equation}
\label{eq:potential}
V(\phi)=m^2 f^2\left[1-\cos\left(\frac{\phi}{f}\right)\right].
\end{equation}

If there is no, or negligible, coupling to the dark gauge fields, then the axion will at late times just oscillate around its quadratic minimum and act as a contribution to cold dark matter. This will not alleviate the Hubble tension. This is why it is essential that we include an appreciable coupling to the gauge field to provide a decay channel. The operator that describes the most leading order interaction is the above dimension 5 operator.
The dimensionful coupling $\M=1/g_{\phi A}$ sets a cut off on the effective theory. This will turn out to be an energy scale many orders of magnitude higher than the relevant scales around matter-radiation equality, so our construction will be a valid effective field theory. It will be useful to express the coupling in terms of $f$ as $g_{\phi A}=\beta/f$. The numerical coefficient $\beta$ is in principle determined by the physics of the dark sector. Of particular importance to fundamental physics will be the specific value of $\beta$ required, which we shall return to. 

In order for the axion field to play a role in alleviating the Hubble tension, one should have $f\sim 0.1\mpl$ and $m\sim H(t_{\text{eq}})$, where the subscript denotes that the quantity is evaluated at the time of matter-radiation equality. This value for $f$ seems quite reasonable from the point of view of fundamental physics. As to the value of $m$, it arises from the breaking of a continuous shift symmetry; in most UV constructions it is exponentially small as it is determined by the strength of instantons. Hence such extremely small values of $m$ are plausible. 

\section{Resonant Decay}
\label{sec:resonant}

We are interested in understanding the resonant decay of the axion field $\phi$ to the gauge field $A^{\mu}$. To this end, we treat $\phi$ as an homogeneous classical background with equation of motion
\begin{align}
\label{eq:axion}
\ddot{\phi}+3H\dot{\phi}+V'(\phi)&=0
\end{align}
On this background, we quantize the four vector potential $A_{\mu}=(A_0,\bf{A})$ (we omit operator symbols in this work). Working in the Coulomb gauge $\bf{\nabla}\cdot \bf{A}=0$ and in the FRW expanding spacetime, the equations of motion of $\bf{A}$ 
are given by
\begin{align}
\label{eq:gaugefield}
\ddot{{\bf{A}}}+H\dot{\bf A}-\frac{{\bf{\nabla}}^2}{a^2} {\bf A}+g_{\phi A} \dot{\phi}~\frac{{\bf \nabla}}{a}\times {\bf{A}}&=0.
\end{align}

Let us first focus on the axion dynamics, as dictated by Eq.~(\ref{eq:axion}). At early times, one has $H\gg m$ and the field is stuck at its initial value, which is naturally $O(f)$. Once $m\gtrsim H$, the axion is released and starts oscillating in its potential, which at large field values exhibits important non-linearities. The homogeneous axion field thus behaves as an underdamped non-linear oscillator, i.e. $\phi(t)=\phi_0 F(t)$, with $F$ being an oscillating function which, in the absence of Hubble friction, is also periodic. As oscillations progress, Hubble friction reduces their amplitudes until the field moves in an effectively quadratic potential and $F(t)\simeq \cos(m t)$.

Let us now study the gauge field dynamics, according to \eqref{eq:gaugefield}. It is useful to rewrite it in Fourier space as follows
\begin{equation}
\label{eq:modes}
\ddot{s}_{\bf{k}, \pm}+H\dot{s}_{\bf{k}, \pm}+\left[\left(\frac{k}{a}\right)^2 \mp \frac{k}{a}\frac{\beta}{f}\dot{\phi}\right]s_{\bf{k}, \pm}=0,
\end{equation}
where $s_{\bf{k}, \pm}$ are the mode functions, $\pm$ denotes the two circular polarizations of the gauge field, and $k\equiv \lvert \bf{k}\rvert$.
Solutions of the equation above can exhibit instabilities, meaning that the mode functions $s_k$ can grow exponentially, according to two distinct but related mechanisms. We can describe the solutions by momentarily neglecting the expansion of the Universe.

First, the effective frequency $\omega^2_k\equiv k(k\mp \beta/f\dot{\phi})$ of a given $k$-mode can turn negative as $\dot{\phi}$ changes sign during the axion oscillations. In the harmonic approximation for the axion, it is easy to see that modes with $k\leq k_{\text{max}}\simeq m\beta \theta_i /2$ are affected by this so-called tachyonic resonance. The time scale of this growth is given by the time that $\dot{\phi}$ remains with a given sign, that is $\delta t\sim 1/m$.

\begin{figure}[t]
   \centering
	\includegraphics[width=0.5\textwidth]{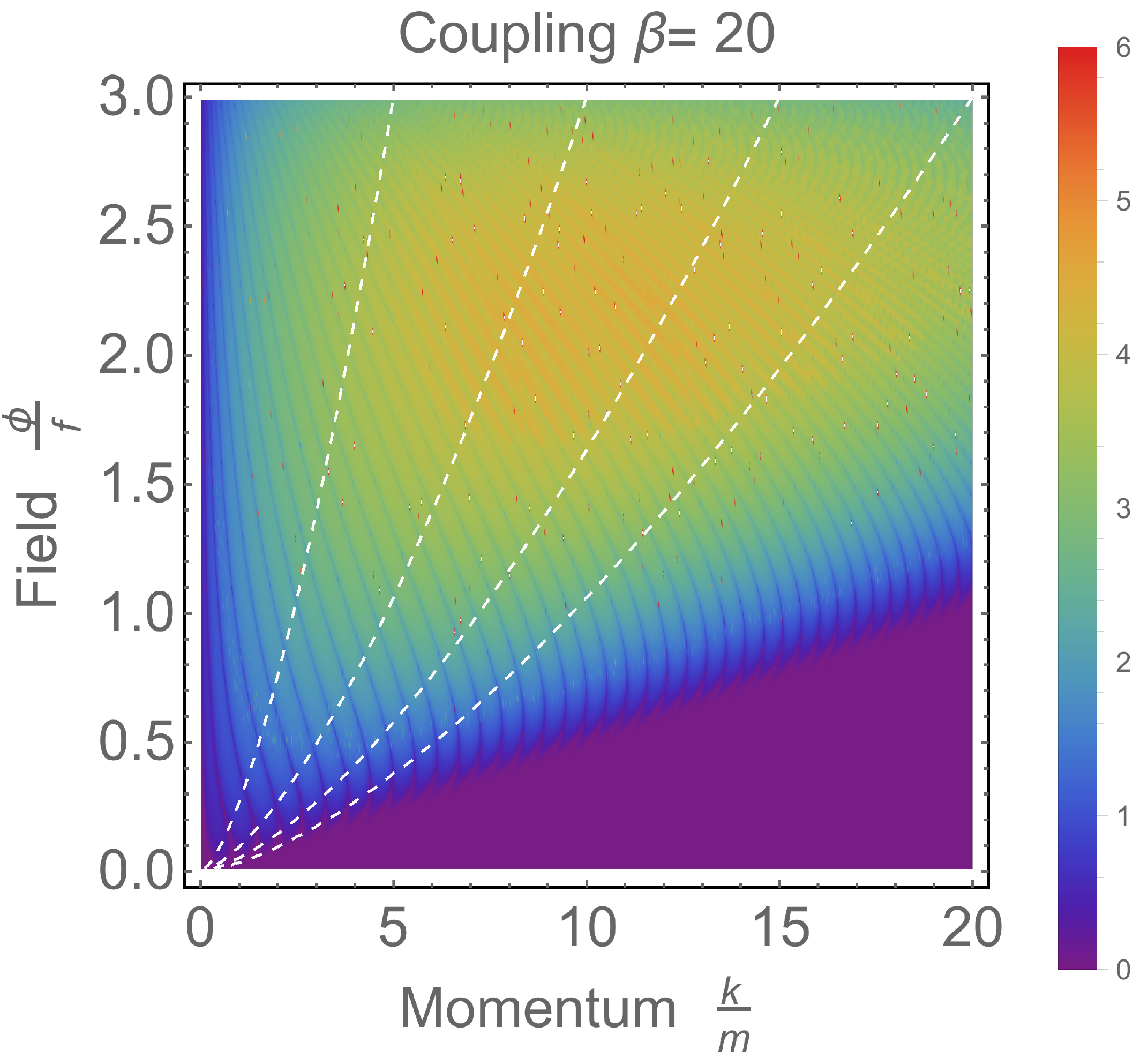}
	\caption{\small Values of the Floquet exponent $\mu_k$ in units of $m$ for \eqref{eq:gaugefield}. We have focused on the $-$ polarization and fixed $\beta=20$ as a representative example. The trajectories of the modes $k=5~m, 10~m, 15~m$ and $20~m$ as the Universe expands are shown as white dashed curves.}
	\label{fig:floquet}
\end{figure}

Secondly, even when the effective frequency is positive, parametric resonance can occur because Eq.~\eqref{eq:modes} takes the form of Hill's equation since $\dot{\phi}$ is a periodic function. Thus, according to Floquet's theorem, the solutions of \eqref{eq:modes} are given by $s_{\bf{k}, \pm}\sim e^{\mu_k t} P(t)$, where $P(t)$ is a periodic function with the same period as the oscillating axion background and $\mu_k$ is in general complex. For certain values of $k$ lying in so-called resonance bands one has $\Re(\mu_k)>0$ and the corresponding mode grows exponentially. In Fig.~\ref{fig:floquet} we show the values of the real part of the Floquet exponent in units of $m$ as a function of $k$ and $\phi_i$, as computed by solving numerically \eqref{eq:axion} and \eqref{eq:gaugefield} (see e.g.~\cite{Amin_2012} for a computational method). One can appreciate that the largest enhancement occurs around $k\simeq \beta m/2, \theta_i\simeq 2$.

If the amplitude of the oscillations is small enough that the axion oscillates in an effectively quadratic potential, \eqref{eq:gaugefield} can be rewritten as a Mathieu equation with resonance bands which coincide with those shown in Fig.~\ref{fig:floquet} for $\theta\lesssim 1$.

Let us now consider the effects of the expansion. Once Hubble friction is included, efficient resonant growth occurs only if its time scale is shorter than the Hubble time, i.e.~if $m>H$ for tachyonic resonance and $\Re(\mu_k)> H$ for parametric resonance. Furthermore, the amplitude of the axion oscillations redshifts as $a^{-3/2}$ and momenta as $a^{-1}$. Therefore, on the one hand, tachyonic resonance is only effective during the first axion oscillations since the range of $k$ modes affected by this resonance decreases with the expansion. On the other hand, a given mode $k$ can now pass through several Floquet bands, as shown in Fig.~\ref{fig:floquet}.

It is qualitatively easy to understand the potential relevance of these dynamics for the Hubble tension. Deep in the radiation era the axion field is overdamped and practically behaves as a dark energy component. Around the epoch of matter-radiation equality, the axion begins to oscillate. Eventually, it would behave as dark matter, since after few oscillations its energy density starts redshifting as $a^{-3}$. By itself, this would actually worsen the goodness of the CMB fit. However, as the axion oscillates it can start dumping its energy into dark gauge fields, as outlined above. The additional energy in the would-be dark matter field can then be rapidly converted into a radiation component. Intuitively, this energy transfer should occur rapidly enough so that the two-field system effectively behaves similarly to the EDE scenario of \cite{Poulin:2018cxd} with $n=2$ and of \cite{Agrawal:2019lmo}, which both significantly alleviate the Hubble tension. 
 
\begin{figure}[t]
   \centering
	\includegraphics[width=0.48\textwidth]{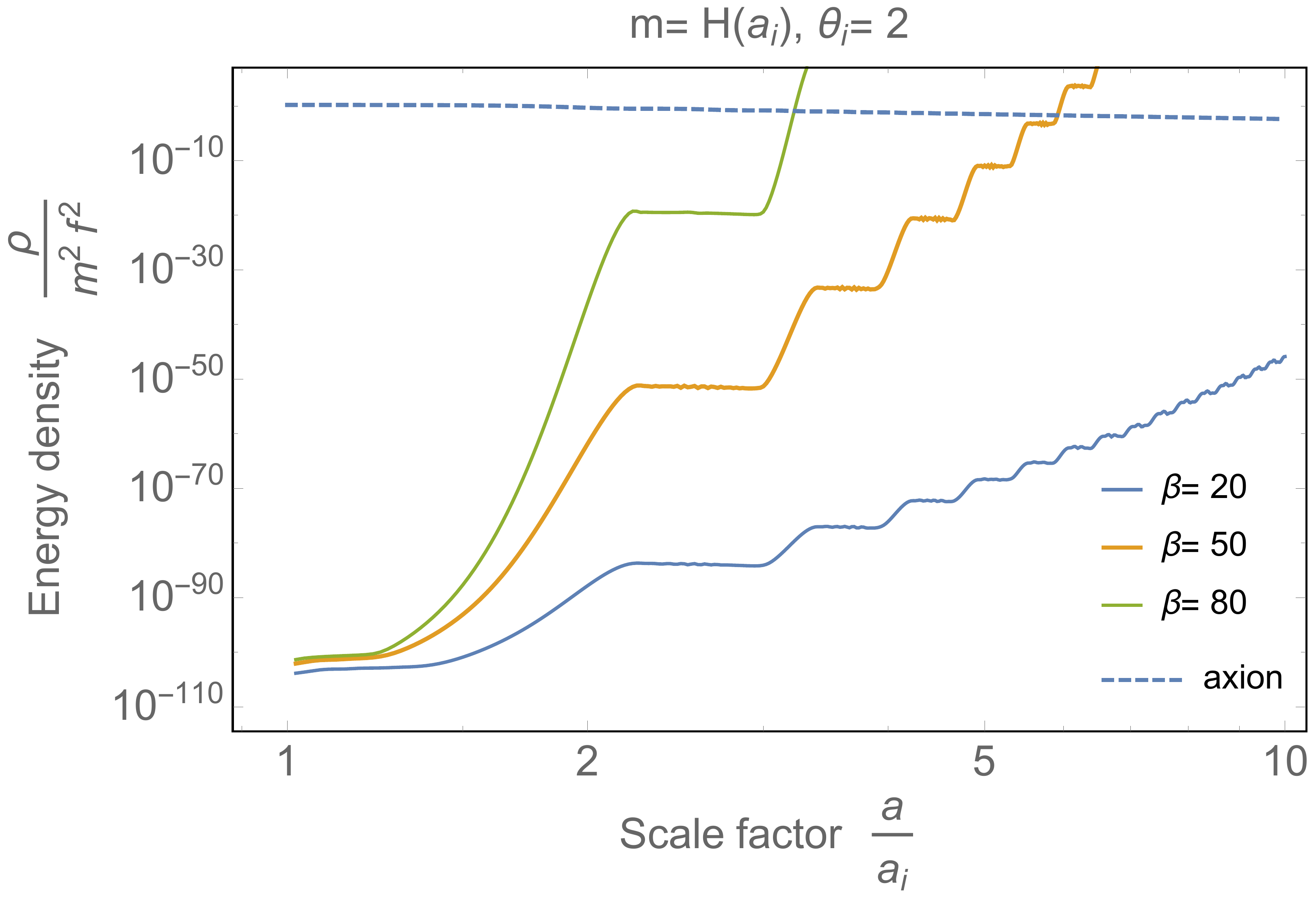}
	\caption{\small Growth of the energy in the $-$ polarization of the gauge field, according to \eqref{eq:gaugefield} and \eqref{eq:engauge} for several values of axion-gauge field coupling $\beta=20, 50, 80$ from left to right in the plot, fixing $\theta_i=2$, assuming a radiation dominated background. The dashed line shows the redshifting energy density of the axion field, for $m=H(a_i)$ and for representative parameters $m=10^{-26}~\text{eV}$ and $f=5\cdot 10^{17}~\text{GeV}$. The calculation has been performed by discretizing the integral in \eqref{eq:gaugefield}, setting a cutoff $k_{\text{cut}}=\beta m a_i $ and sampling the energy density over $10\beta$ points. We have checked that the results are not significantly affected by increasing the number of points and/or the cutoff.}
	\label{fig:growth}
\end{figure}

The question of how fast the energy transfer should occur for the mechanism to alleviate the Hubble tension can be quantitatively addressed only by performing a dedicated fit to cosmological data, as we will do in the next section. Here instead we provide evidence that in principle the energy in the gauge fields can be sufficiently amplified during the first few axion oscillations. 
We solve the system of equations \eqref{eq:axion} and \eqref{eq:gaugefield} numerically, assuming a radiation-dominated background to fix ideas and $m=H$, $\theta_i=2, \dot{\theta}_i=0$ as initial conditions for the axion field, while the gauge fields are initialized according to the Bunch-Davies vacuum. We then compute the energy density in the gauge fields as 
\begin{equation}
\label{eq:engauge}
\rho_A=\frac{1}{2a^4}\int \frac{d^3 k}{(2\pi)^3}\sum_{\pm}\left(a^2\lvert\dot{A}_\pm\rvert^2+k^2\lvert A_{\pm} \rvert^2 -2k\right),
\end{equation}
which we plot in Fig.~\ref{fig:growth} for several values of $\beta$, together with the axion energy density. Notice that the initial energy density in the gauge fields is extremely small compared to the axion energy density, since $\rho_A/V\simeq (k^4/m^2 f^2)\sim (m/f)^2\simeq 10^{-110}$ for representative values $m\sim 10^{-26}~\text{eV}$ and $f\sim 10^{17}~\text{GeV}$ (which roughly correspond to the bestfit values which we determine in Sec.~\label{sec:fit}). The ``steps'' in Fig.~\ref{fig:growth} correspond to the oscillations of the axion field. Therefore, one can see that for $\beta\gtrsim 50$ the energy density in the gauge fields becomes comparable to the background axion energy density after the axion completes five/six oscillations. For even larger $\beta$, sufficient growth occurs after only few oscillations (and even after just the first oscillation for $\beta\gtrsim 80$). For $\theta_i>2$, the resonant enhancement is faster and smaller values of $\beta$ are sufficient to achieve the required rapid growth. This can be understood by looking at Fig.~\ref{fig:floquet}, since the largest parametric enhancement occurs around $\theta\sim 2$.

Of course our linear analysis breaks down before the energy density of the gauge fields becomes comparable to the axion energy density shown in Fig.~\ref{fig:growth}. Nonetheless, the relevant time scale needed to achieve the energy transfer should still be correctly captured by our simple estimate. 

An important caveat is now in order: the analysis above further neglects the backreaction of the gauge fields on the axion field, which causes a growth of axion fluctuations and can thus further alter the axion-gauge field dynamics. Dedicated lattice calculations are required to realistically assess the implications of such backreaction, as well as the detailed energy transfer from the axion to the gauge field. These have been performed in \cite{Kitajima:2017peg} for the case in which $\phi$ is the QCD axion. Very interestingly, the results of~\cite{Kitajima:2017peg} show that the energy transfer can indeed efficiently occur in the first few oscillations, for the same range of $\beta$ shown in Fig.~\ref{fig:growth}. However, when the growth is so strong that the energy density in the gauge field is comparable to that in the axion field already during the first half-oscillation, then backreaction serves as an effective friction on the motion of the rolling axion, whose relic abundance is then enhanced rather than diluted. Following~\cite{Kitajima:2017peg}, the upper bound on $\beta$ to avoid such a situation can be estimated as follows. Focusing on the energy in the $k_{\text{max}}$ mode, after the first half oscillation, tachyonic resonance gives $\rho_{A, k}\sim (k_{\text{max}}/a_i)^4~\exp\left(2\int dt ~k_{\text{max}}/a \right)\sim (\beta m \theta_i/2)^4 \exp\left(\beta \theta_i \right) $. By imposing that the latter energy is smaller than the axion energy density at the end of the first half oscillation, i.e. $1/2 m^2 f^2\theta_i^2$, one finds 
\begin{equation}
\label{eq:backreaction}
\beta\lesssim 2\theta_i^{-1}\log\left(\theta_i\frac{f}{m}\right)\sim 100-200
\end{equation}
for $\theta_i\sim 1-2$ and $m\sim 10^{-26}~\text{eV}, f\sim 5\cdot 10^{17}~\text{GeV}$. We conclude that the relevant range of axion-gauge field coupling required for this mechanism to alleviate the Hubble tension is $50 \lesssim \beta\lesssim 150$. 

While the above model is a perfectly sensible effective field theory, as this operator respects the shift-symmetry of the axion, this is nevertheless an uncomfortably large value for the axion-photon coupling, and begs the question whether it makes sense from the point of view of UV physics.

\section{Other Models}
\label{sec:other}

This issue can be addressed in the context of a broader class of models. The discussion above has focused on an axion field decaying to Abelian gauge fields. Similar dynamics can be obtained also with scalar fields (see also \cite{Kaloper:2019lpl}). For instance, one can consider a real scalar field $\phi$ that is coupled to an essentially massless secondary scalar field $\chi$ with Lagrangian
\begin{equation}
\mathcal{L}={1\over2}(\partial\phi)^2+{1\over2}(\partial\chi)^2-{1\over2}m^2\phi^2 -{1\over2}\varepsilon\,\phi\,\chi^2
\end{equation}
The cubic interaction also allows for tachyonic and parametric resonance. One can show that efficient energy transfer from $\phi$ to $\chi$ requires
\begin{equation}
\varepsilon \gg {m^2\over\phi_i}
\label{RETSS}\end{equation}
where $\phi_i$ is the amplitude of the scalar field at the onset of oscillations $H\sim m$. Now since we are considering extremely small masses for $m\gtrsim 10^{-28}$\,eV, this condition on $\varepsilon$ seems very easy to satisfy. So again from the effective field theory point of view this is perfectly reasonable (one can further check that loop corrections to the mass of $\phi$ or $\chi$ are negligible too, unless $\varepsilon$ is taken many orders of magnitude higher than this bound). 

However, one can also wonder about the UV completion of this model. Since the theory is already dimension 4 this does not seem an issue. On the other hand, one might wish to justify the masslessness of $\chi$, especially since the above Lagrangian does {\em not} have a shift symmetry for $\chi$. This can be done as follows: Consider a complex scalar $\Phi$, with a spontaneously broken global $U(1)$ symmetry (such a symmetry can be broken non-perturbatively and/or by quantum gravity, but we assume such corrections are small in this discussion)
\begin{equation}
\mathcal{L}=|\partial\Phi|^2-\lambda(|\Phi|^2-v^2)^2
\end{equation}
Here there can be resonant decay to the Goldstone mode (see e.g.,~\cite{Kasuya:1997ha}). Expanding around the VEV $v$ as $\Phi = v+(\phi+i\chi)/\sqrt{2}$, the mass of the radial mode $\phi$ is $m=\sqrt{2\lambda}\,v$ and the cubic coupling to the Goldstone mode $\chi$ is $\varepsilon= 2\lambda v$. Hence the condition for rapid energy transfer (\ref{RETSS}) is $|\phi_i|\gg v$. This means that the radial mode should initially sit high up in its quartic potential, far from the VEV and puts one in a regime in which the standard resonance analysis does not apply. This may be considered analogous to the requirement $|\phi_i|\sim f \gg \Lambda= f/\beta$ (i.e., $\beta\gg 1$) in the axion-gauge field system studied earlier.

Finally, non-resonant energy transfers may also be envisioned. For example, Ref.~\cite{Berghaus:2019cls} considers a scenario with an axion decaying to non-Abelian gauge fields through sphaleron effects. However, the condition of rapid enough decay may be challenging to achieve.
Or one could simply consider ordinary perturbative decays by considering much larger couplings. For example, one could just allow $\phi$ to decay into massless (dark) fermions $\psi$ through the Yukawa interaction $\Delta\mathcal{L}=y\phi\bar{\psi}\psi$. The  perturbative decay rate is
\begin{equation}
\Gamma_{pert} = {y^2\,m\over 8\pi}
\end{equation}
So in order for the scalar $\phi$ to decay shortly after it starts rolling at $H\sim m$, one needs $\Gamma_{pert}$ to be not too much smaller than $m$. This means the Yukawa coupling should be $y=\mathcal{O}(1)$.  This represents a large breaking of any shift symmetry for $\phi$. 
Therefore, such a set-up does not explain why $\phi$ is so light. While one can simply set the renormalized mass of $\phi$ to its desired value $m\sim H_{eq}$ and since the model does not have any heavy degrees of freedom, this is again not a direct problem with the effective field theory. But it also begs the question as to whether it has a UV embedding.

\section{Effective fluid description}
\label{sec:fluid}

In order to quantitatively assess the relevance of this scenario for the Hubble tension, a numerical realization in a Boltzmann code is required. 
While this is a relatively straightforward task for single field models, realizing the coupled axion-gauge field dynamics above is challenging with existing Boltzmann codes. Therefore, in this work we rather make a first step towards a fully realistic numerical analysis by employing an effective fluid description of the axion-gauge field dynamics, which captures some of its most important features. This analysis is meant to be valid in a coarse-grained sense, meaning after averaging over an appropriate time scale, for instance over one period for the oscillating axion field (see~\cite{Poulin:2018cxd, Lin:2019qug} for similar approaches in the context of the Hubble tension). Our simplification will proceed in two steps: first, we will introduce an effective two-fluid description of the axion-gauge field system, in which we encode the energy transfer from the axion fluid to the gauge field fluid. Secondly, we will further reduce to a single fluid system. The latter is the description which we implemented in a Boltzmann code. While our approach may not be entirely justified based on the complex two-field dynamics, we think that it serves the purpose of providing an understanding of whether the mechanism can help in alleviating the Hubble tension. 

Let us first focus on the background dynamics of the axion-gauge field system. As shown in \cite{Poulin:2018dzj}, at the background level a homogeneous scalar field with potential \eqref{eq:potential} can be effectively described as a fluid with equation of state parameter
\begin{equation}
\label{eq:wphi}
w_{\phi}(a)=-1+\frac{1}{1+(a_c/a)^3},
\end{equation}
where $a_c$ is the scale factor evaluated at the time when the scalar field starts rolling down its potential. On the other hand, massless gauge fields behave as radiation and thus have an effective equation of state parameter $w_A=1/3$. In order to capture the energy transfer from the axion to the gauge field, we can then make use of the following effective coupled fluid equations:
\begin{align}
\label{eq:2fl}
\dot{\rho}_{\phi}+3H(1+w_\phi)\rho_\phi&=-\Gamma(t)\rho_\phi\\
\dot{\rho}_{A}+4H\rho_A&=\Gamma(t)\rho_\phi,
\end{align}
where $\rho_{\phi, A}$ are the energy densities of the axion and gauge field fluids and $\Gamma(t)$ is an effective decay rate. The latter can in principle be numerically extracted by means of a lattice simulation (such as the one presented in~\cite{Kitajima:2017peg} in the context of the QCD axion coupling to hidden photons). Such an analysis goes beyond the aim of this paper. However, in a rather model-independent fashion, we can parametrize the decay rate as follows
\begin{equation}
\label{eq:rate}
\Gamma(t)=\frac{1}{1+\gamma(a_r/a(t))},
\end{equation}
where $\gamma(x)$ is a function such that $\gamma(x)\gg 1$ for $x\gg 1$ and $\gamma(x)\ll 1$ for $x\ll 1$. To set ideas, we consider here $\gamma(x)=x^p$, with $p>0$. Here $a_r$ is the scale factor evaluated at the time $t_r$ at which parametric resonance becomes efficient. Obviously $a_r>a_c$; its precise value depends on the coupling strength $g_{\phi A}$, and in particular on the value of $\beta$. The regime of our interest is $\beta\gg 1$ and in this case the energy transfer from the homogeneous axion background to the gauge field can be extremely efficient, as discussed in the previous section, in such a way that the axion field dumps all its energy during at most the first few oscillations. To reflect this feature in our parameterization, we will consider $a_r\gtrsim a_c$ and $p\gg 1$.

\begin{figure}[t]
   \centering
	\includegraphics[width=\columnwidth]{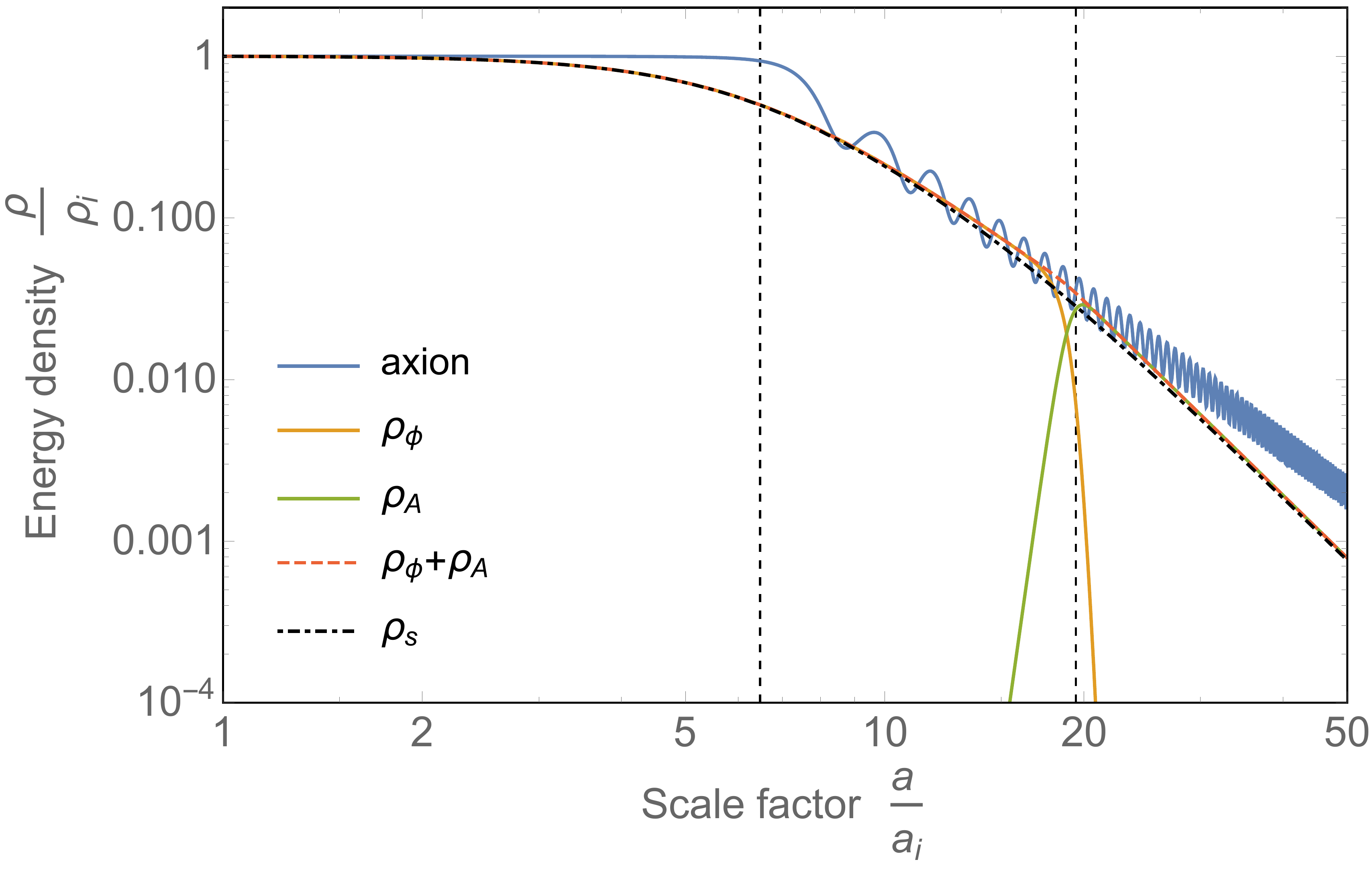}
	\caption{\small Comparison between: the axion energy density computed from the field theory model (solid blue), the axion-gauge field energy density according to the two-fluid approximation \eqref{eq:2fl} (solid orange, green, red) and according to the single fluid approximation with equation of state parameter \eqref{eq:wphiA}. In order to produce this plot, we have chosen as an example $H_i=3\,m,\, \theta_i=3, \, a_c\simeq 6.5\, a_i,\, a_r= 3\, a_c, \, a_d=0.9\, a_r$.}
	\label{fig:enfluid}
\end{figure}

We now proceed to the second step of our simplification. Namely, we consider a single fluid approach to describe the two-fluid system presented above. 
We start by considering the following equation of state parameter
\begin{equation}
\label{eq:wphiA}
w_{s}(a)=-1+\frac{1}{1+(a_c/a)^3}+\frac{1/3}{1+(a_d/a)^d},
\end{equation}
with $d>0$ and $a_d>a_c$. A single fluid with the equation of state parameter above features two transitions: the first one from $-1$ to $0$ around $a_c$, the second one from $0$ to $1/3$ around $a_d$. This matches the two-fluid picture, if $a_d\simeq a_r$.
We compare the evolution of the energy densities of the two-fluid system, according to \eqref{eq:2fl}, \eqref{eq:wphi}, \eqref{eq:rate}, with the energy density $\rho_s$ of the single fluid with equation of state parameter \eqref{eq:wphiA} in Fig.~\ref{fig:enfluid} for an example choice of parameters. For simplicity, we work in a radiation-dominated background, although our conclusions are not affected by the background evolution. Furthermore, we fix $p=30$ and $d=4$ as a representative example and choose $a_d\simeq a_r$. We observe a very good agreement between the two effective descriptions of the evolving axion and gauge field system. The agreement can be made even more precise considering larger values of $d$. However, we have verified that the results of the fit to cosmological data are not significantly affected by the choice of $d$, as long as $d\geq 4$. Notice that the single-fluid parametrization of the axion-gauge field system, according to~\eqref{eq:wphiA}, is not expected to provide an accurate description of the background evolution at late times. Indeed, even though the axion field transfers a significant fraction of its energy density to gauge fields at early times, a small fraction of its energy density likely remains in the form of axions and redshifts like matter \cite{Kitajima:2017peg}. At late times, significantly after recombination, this small fraction is expected to dominate over the energy density in the gauge fields, since the latter redshifts faster. Therefore, at late times the axion-gauge field system should be dominated by a dark matter component, rather than by radiation as described by \eqref{eq:wphiA}. Nevertheless, at such late epochs the contribution of our dark sector to the total energy density is very small, thus we expect our parametrization \eqref{eq:wphiA} to be valid to the aim of assessing the Hubble tension in our scenario.

The validity of an effective fluid description at the level of perturbations is more subtle. In particular, the sound speed of the effective fluid perturbations should exhibit a dependence on the scale factor as well as on the momentum $k$ of the given perturbation mode (see~\cite{Poulin:2018dzj} for a detailed discussion in the context of a single oscillating axion field). In this work we will consider a sound speed which tracks the behavior of the equation of state parameter $w_{\phi A}$, i.e.
\begin{equation}
\label{eq:cs2}
c_s^2=1-\frac{1}{1+(a_c/a)^3}+\frac{1/3}{1+(a_d/a)^d}.
\end{equation}
Additionally, in order to understand how a different treatment of perturbations may affect our results, we will briefly consider fixed values of the sound speed $c_s^2$, in between the asymptotic early and late time values $c_s^2=1$ and $c_s^2=1/3$.
An important caveat is in order before moving to the next section: the realistic two-field model may exhibit a peculiar behavior for perturbations which is not captured by our simple effective fluid approach. Therefore, our analysis should be considered purely as a first step towards a more realistic numerical implementation, which is left for future work. 
Finally, let us also remark that our parameterization \eqref{eq:wphiA} can capture a broad range of models where dark energy decays to a radiation-like species, beyond the axion-gauge field system. As such, the results presented in the next section may offer insight for such models as well.

\section{Datasets and Results}
\label{sec:fit}

We are now ready to present the results of a numerical fit of our single fluid parameterization of the decaying ultralight scalar (dULS) model to cosmological datasets. 
We consider the following data sets: Planck 2018 high-$\ell$ and low-$\ell$ TT, TE, EE and lensing data~\cite{Aghanim:2019ame} (from now on Planck18); BAO measurements from 6dFGS at $z = 0.106$~\cite{Beutler:2011hx}, from the MGS galaxy sample of SDSS at $z = 0.15$~\cite{Ross:2014qpa} (BAO smallz), from the CMASS and LOWZ galaxy samples of BOSS DR12 at $z = 0.38$, $0.51$, and $0.61$~\cite{Alam:2016hwk} (bao DR12); the Pantheon Supernovae data sample~\cite{Scolnic:2017caz} (Pantheon) and the latest measurement of the present day Hubble rate from the SH$_{0}$ES program $H_0 = 74.03\pm 1.42$ km/s/Mpc~\cite{Riess:2019cxk}. 
We implement our fluid model (which we refer to as dULS, for decaying Ultra-Light Scalar) in the Boltzmann code {\tt{CLASS}}~\cite{Lesgourgues:2011re, Blas:2011rf}, by making use of its dark energy fluid section. Furthermore, we include the possibility of a varying sound speed in the perturbations part of the code. We model neutrinos as two massless and one massive species with $m_{\nu}= 0.06~\text{eV}$, following the Planck collaboration.

Our construction features three additional parameters with respect to $\Lambda$CDM: the relic abundance of the fluid with equation of state parameter given in \eqref{eq:wphiA}, $\Omega_{\text{dULS}}$; the value of the scale factor $a_c$ in \eqref{eq:wphiA}; a parameter $g_d\equiv a_d/a_c$ which quantifies the time scale of the transition from a CDM-like fluid to radiation., with the obvious requirement $g_d\geq 1$. In terms of the original axion-gauge field model, the larger $g_d$, the larger the number of oscillations of the axion field before it undergoes strong resonant decay. For simplicity, we actually present here results only for a fixed representative value $g_d=1.1$. Therefore, the fitted model only has two varying extra parameters beyond $\Lambda$CDM. In terms of the original axion-gauge field model, this choice implies a very rapid energy transfer from the axion field to the gauge field, which occurs during the first axion oscillation. As mentioned in Sec.~\ref{sec:resonant} above \eqref{eq:backreaction}, the feasibility of such a rapid energy transfer probably deserves a dedicated numerical study which includes backreaction effects.\footnote{We have also performed runs with $g_d$ free to vary and found results which are very similar to the ones presented here. Furthermore, we find that the range $g_d\gg 2$ is disfavored by the data compared to $g_d < 2$, although for $g_d \gtrsim 2$ we find that this model still alleviates the Hubble tension and has a smaller $\chi^2$ than $\Lambda$CDM with the same dataset used in this work.} In the dULS model, the sound speed of fluid perturbations is given by \eqref{eq:cs2}. 

Furthermore, we study the dependence of our results on the sound speed by considering variants of the model with a fixed value of the sound speed $c_s^2\geq 1/3$ and with varying $g_d$. We refer to these variants as dULS$_{c_s^2}$ and we will briefly mention their potential relevance to assess the Hubble tension in our framework.

\begin{figure*}[t]
   \centering
	\includegraphics[width=\textwidth]{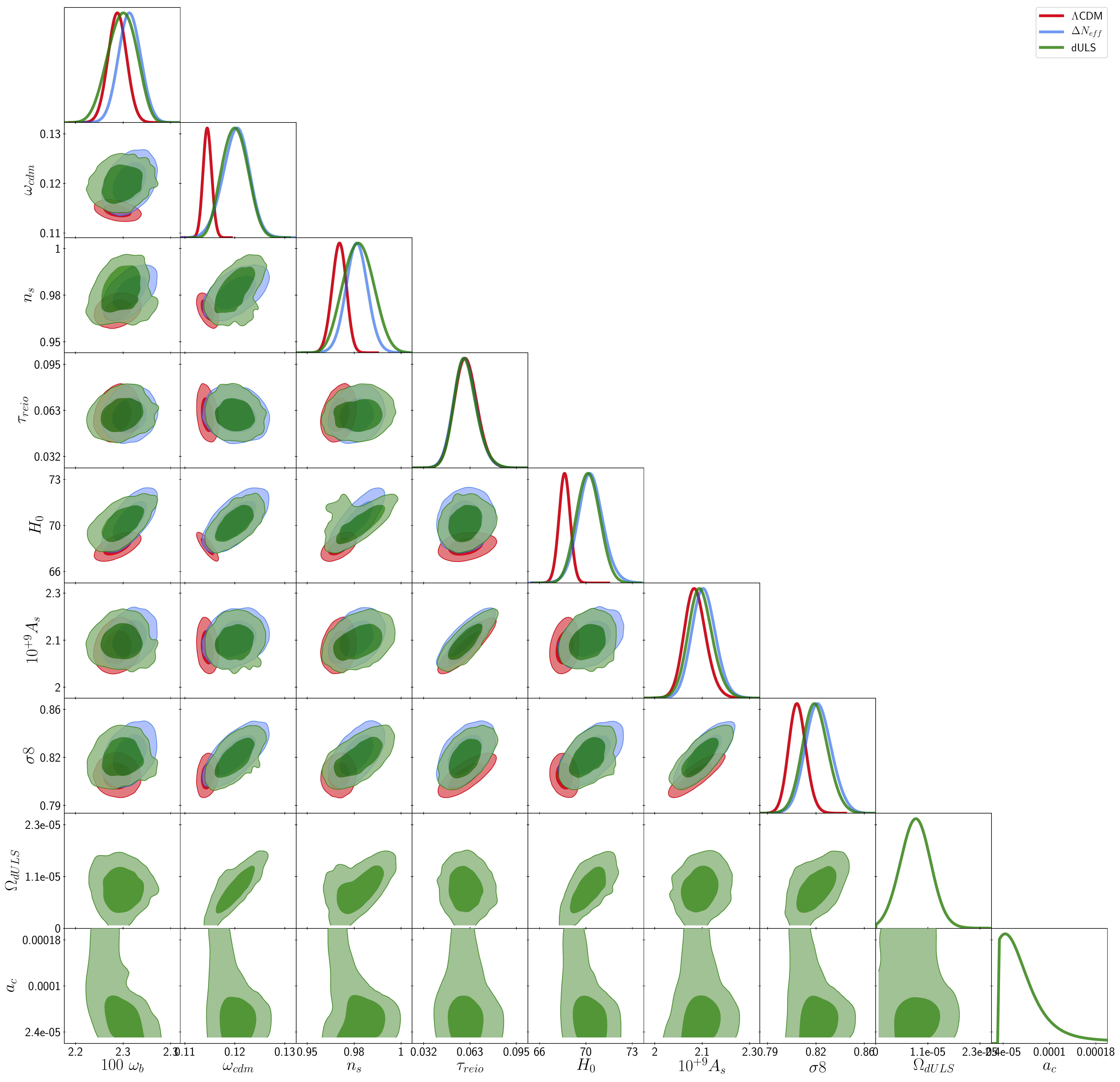}
	\caption{\small Posterior distributions for cosmological parameters in the dULS model (green), compared to $\Lambda$CDM (red) and the $\Delta N_{eff}$ (blue) model. We used the joint Planck 18 + BAO + Pantheon + SH$_{0}$ES likelihood (see beginning of Sec.~\ref{sec:fit}).}
	\label{fig:dEDE}
\end{figure*}

We use {\tt Monte Python}~\cite{Audren:2012wb, Brinckmann:2018cvx}, in its version $3.3.2$, to perform a Markov chain Monte Carlo investigation of the model.\footnote{For all of the analyses presented in this paper, the Gelman-Rubin convergence criterion~\cite{Gelman:1992zz} is respected, since $R-1< 0.1$. In particular, for the $\Lambda$CDM and $\Delta N_{\text{eff}}$ models we have $R-1 < 0.005$.} For comparison, we also obtain chains for the $\Lambda$CDM and $\Delta N_{\text{eff}}$ models. The results for cosmological parameters are reported in Table~\ref{table:results} and in Fig.~\ref{fig:dEDE}. 

\begin{figure}[t]
   \centering
	\includegraphics[width=\columnwidth]{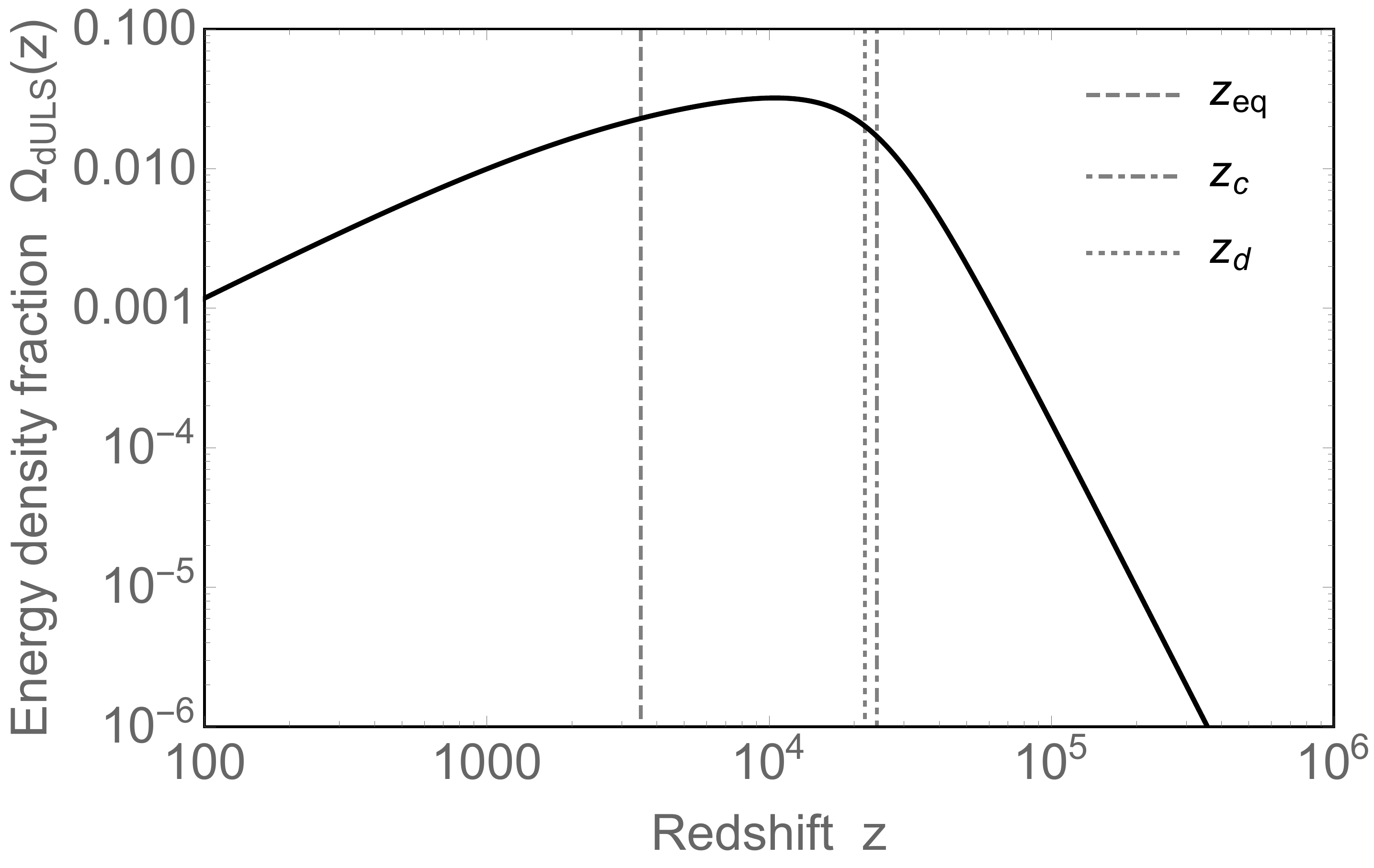}
	\caption{\small The energy density fraction $\Omega_{\mbox{\tiny{dULS}}}=\rho_{\mbox{\tiny{dULS}}}/\rho_{tot}$ of the single fluid with equation of state \eqref{eq:wphiA} as a function of redshift $z$. The vertical lines indicate the redshift of matter-radiation equality $z_{eq}$, the redshift $z_c$ when the equation of state parameter starts deviating from $-1$, and the redshift $z_d$ after which the fluid redshifts as radiation.}
	\label{fig:energyvsz}
\end{figure}

Let us first discuss results for the dULS model. The behavior of the energy density fraction in the effective fluid for the bestfit values of parameters in Table~\ref{table:results} is reported in Fig.~\ref{fig:energyvsz}. One can appreciate that the data prefers a transition from dark energy to matter around $z_c\simeq 24,000$, significantly before recombination. Taking into account our assumption $g_d=1.1$, the transition to a radiation-like fluid should occur around $z_d\simeq 22~000$. The actual contribution to the total energy density is $\sim 3-4\%$ slightly after $z_c$. Therefore, the model is expected to alleviate the Hubble tension, without being able to fully solve it. 
Indeed we find $H_0=69.9_{-0.86}^{+0.84}~\text{km/s/Mpc}$ with $H_0=69.92~\text{km/s/Mpc}$ being the best fit value. The dULS model can be compared to the $\Delta N_{\text{eff}}$ model, as reported in Table~\ref{table:results} and in Fig.~\ref{fig:dEDE}. 
The dULS model is preferred with respect to $\Lambda$CDM model by $\Delta \chi^2\approx -9$, whereas the $\Delta N_{\text{eff}}$ model is preferred only at $\Delta \chi^2\approx -3$.\footnote{Several criteria exist to compare the performance of two models with different number of parameters. For instance, the Akaike criterion (see e.g.~\cite{Akaike, Liddle:2007fy}) applied to the dULS and $\Lambda$CDM models gives: $\Delta\text{AIC}\equiv \Delta\chi^2-2\Delta n\approx -5$ if we only consider the two extra free parameters used in our run and $\Delta\text{AIC}\approx - 3$ if we also count $g_d$, with $\Delta n$ being the difference in the number of parameters between the two models. The same criterion applied to the $\Delta N_{\text{eff}}$ model gives  instead $\Delta\text{AIC}\approx-1$. Thus, according to this criterion, the dULS model performs better than the $\Delta N_{\text{eff}}$ model.} As seen in Table~\ref{chi2}, which shows the individual contributions to $\chi^2$, the improvement in $\Delta\chi^2$ of the dULS model with respect to the $\Lambda$CDM model is driven by the fit to SH$_0$ES, as expected. The improvement with respect to the $\Delta N_{\text{eff}}$ model is instead due to a combination of a better fit to SH$_0$ES as well as Planck high-l and low-l TT data.

\begin{table*}
  \begin{tabular}{|l|c|c|c|}
    \hline\hline
    Parameter &~~$\Lambda$CDM~~&~~~dULS~~~&~~~$\Delta N_{\text{eff}}$\\ \hline \hline
    $100~\omega_b$ & $2.254~(2.26)_{-0.014}^{+0.013}$ & $2.26~(2.264)_{-0.019}^{+0.021}$ & $ 2.272 (2.262)$ \small${}^{+0.016}_{-0.016}$ \\
    $\omega_{\rm cdm}$&$0.1183~(0.1189)_{-0.00092}^{+0.00087}$& $0.1239~(0.124)_{-0.0026}^{+0.0026}$   &
    $0.124 (0.1231)$ \small${}_{-0.0028}^{+0.0027}$\\
    $10^{9}A_s$& $2.122~(2.123)_{-0.035}^{+0.03}$ &$2.137~(2.134)_{-0.037}^{+0.034}$ & $2.147 (2.131)$ \small${}_{-0.036}^{+0.033}$ \\
    $n_s$& $0.97~(0.9699)_{-0.0036}^{+0.0038}$ &$0.9802~(0.9857)_{-0.0082}^{+0.0086}$ & $0.9792 (0.9779)$ \small${}_{-0.0059}^{+0.0058}$  \\
    $\tau_{\rm reio}$ &  $0.06053~(0.06027)_{-0.0084}^{+0.007}$& $0.06049~(0.06011)_{-0.0081}^{+0.007}$ & $0.06018 (0.05912)$ \small${}_{-0.0083}^{+0.0072}$\\
    $H_0$  & $68.24~(68.06)\pm 0.41$ & $69.9~(69.92)_{-0.86}^{+0.84}$ &  $70.08 (69.64)$  \small${}_{-0.95}^{+0.91}$  \\
     $10^{6}\Omega_{\text{dULS}}/\Delta N_{\text{eff}}$  &$-$  & $8.764~(8.278)_{-3.5}^{+3.1}$ & $0.3401 (0.2825)$ \small${}_{-0.16}^{+0.15}$ \\
    $10^5 a_c$& $-$  &$5.988~(4.159)_{-4.4}^{+1.1}$&   $-$\\
    $g_d$& $-$  &$\text{fixed to}~1.1$  &   $-$\\
   \hline
    $\sigma_8$ & $0.8097~(0.8119)_{-0.0067}^{+0.0061}$&$0.8226~(0.8243)_{-0.0097}^{+0.0091}$ & $0.825(0.821)$ \small${}_{-0.0095}^{+0.0095} $\\
    \hline
  \end{tabular}
  \caption{The mean (best-fit in parenthesis) $\pm1\sigma$ error of the cosmological parameters obtained by fitting $\Lambda$CDM, the dULS and the $\Delta N_{\text{eff}}$ models to our combined cosmological dataset.}
  \label{table:results}
\end{table*}

\begin{table}[h!]
\centering
{\renewcommand{\arraystretch}{1.25} 
{\begin{tabular}{|c | c | c | c |}
\hline
Dataset & $\Lambda$CDM & dULS & $\Delta N_{\text{eff}}$ \\
\hline
 Planck highl TTTEEE  & 2352.18 & 2353.94  & 2357.80  \\
\hline
  Planck lowl EE   & 397.44 & 397.18  & 397.04\\
\hline
  Planck lowl TT  & 22.71 & 20.88 & 21.78\\
\hline
 Planck lensing  & 8.84 & 9.68  & 9.41\\
\hline
 Pantheon & 1027.06 & 1026.92  & 1026.92\\
\hline
  SH$_0$ES 2019 & 17.71 & 8.39  & 9.56\\
\hline
 bao boss dr12 & 3.81 & 3.36  &  3.46\\ 
\hline
  bao smallz 2014 & 1.45 & 1.94  & 1.80\\
\hline
Total  & 3831.19 & 3822.28  &  3827.78\\
\hline
  $\Delta \chi^2$ & $0$&$-8.91$ & $-3.41$\\
\hline
\end{tabular}}
}
\caption{\small Contributions to the total $\chi^2_{\rm eff}$ for individual datasets, for the best-fits of $\Lambda$CDM, dULS and $\Delta N_{\rm eff}$ models.  \label{chi2}}
\end{table}

In spite of their power in alleviating the Hubble tension, both the dULS model and the  $\Delta N_{\text{eff}}$ model exhibit larger values of the $\sigma_8$ and $\omega_{\text{cdm}}$ parameters with respect to $\Lambda$CDM. Therefore, the $S_8=\sigma_8(\Omega_m/0.3)^{1/2}$ tension between CMB and cosmic shear measurements (e.g., a recent joint analysis of KIDS1000+BOSS+2dfLenS~\cite{Heymans:2020gsg} finds a smaller values of $S_8$, with a $\sim 3\sigma$ discrepancy with \emph{Planck} data assuming the $\Lambda$CDM model) is exacerbated in the dULS and $\Delta N_{\text{eff}}$ models with the cosmological dataset considered in this work. This is a feature common to all EDE-like solutions of the Hubble tension. Indeed these solutions add a certain amount of energy density in the form of radiation (which may or may not decay faster than radiation after equality). In order to keep the CMB angular scale at equality fixed, the dark matter energy density also has to be increased and this leads to an enhancement of the amplitude of the matter power spectrum at late times compared to the $\Lambda$CDM model. Furthermore, the fit may be worsened with the inclusion of other data sets, including large scale structure (see e.g.~\cite{Hill:2020osr,Ivanov:2020ril,DAmico:2020ods} and~\cite{Chudaykin:2020acu} for recent contrasting takes on this issue and its relation to the Hubble tension).

Let us now translate our results from the language of the effective fluid model to that of the ultralight scalar theory. We do so for the bestfit values of parameters presented in Table~\ref{table:results}. First, the axion mass can be roughly determined by setting $m= \alpha H_c$, where by means of {\tt{CLASS}} we extract $H_c\simeq 5\cdot 10^{-27}~\text{eV}$. The value of $\alpha$ depends on the initial axion field value and can be estimated as $\alpha\simeq \sqrt{V'(\phi_i)/\phi_i}/m$. For $\phi_i\gtrsim 2 f$, one finds $\alpha\gtrsim 1.5$, with $\alpha\simeq 4$ when $\phi_i\gtrsim 3 f$. Therefore, we find $m\sim 10^{-26}~\text{eV}$. 
Second, the axion decay constant can be roughly determined by matching the initial energy density fraction shown in Fig.~\ref{fig:dEDE}, which we find to be $\rho_{\mbox{\tiny{dULS}}}(z\simeq z_{c})\simeq 20~\text{eV}^4$, to $V\sim m^2 f^2$. We find $f\simeq 5\cdot 10^{17}~\text{GeV}$.
Finally, the value of the axion-gauge field coupling $\beta$ can be roughly estimated by noticing that $a_d=1.1 a_c$ implies that the energy transfer should efficiently occur during the first oscillation of the axion field. According to Fig.~\ref{fig:growth}, this can be achieved for $\beta\gtrsim 80$. As mentioned above, this regime deserves further numerical investigation to assess the relevance of backreaction effects. Our investigation reveals that slightly smaller values of $\beta$ should still alleviate the Hubble tension, although quantitatively the improvement over the $\Lambda$CDM model in terms of $\Delta \chi^2$ may be weakened.

Let us now briefly discuss the dULS$_{c_s^2}$ model, without presenting detailed results for cosmological parameters. We consider two choices $c_s^2=0.5$ and $c_s^2=1/3$. We find that both variations of our fluid model prefer similar values of $a_c$ as in the dULS model. However, a slower transition to radiation is also preferred, with $z_d\simeq (1/3)z_c\simeq 5000$. This allows for several oscillations of the axion field before the latter actually decays to radiation, thereby smaller values of $\beta\lesssim 70$ are preferred. When $c_s^2=0.5$, the Hubble parameter turns out to be slightly larger than in the dULS model, with $H_0=70.17_{-0.69}^{+0.9}~\text{km/s/Mpc}$. Most importantly, the goodness of the fit with respect to $\Lambda$CDM is further improved to $\Delta \chi^2=-11$, with three extra parameters. For $c_s^2=1/3$ instead $H_0$ is very similar to the dULS case, and the $\Delta \chi^2$ is somewhat worse. 
While the results for $c_s^2=0.5$ are encouraging, the simplified parameterization of fluid perturbations in the dULS$_{c_s^2}$ variants is not expected to be realistic, since in general the sound speed will vary with time. Nonetheless, results for the dULS$_{c_s^2}$ model show that the goodness of the fit can be affected by the specific treatment of perturbations (as also found in \cite{Poulin:2018cxd, Lin:2019qug}). A more realistic implementation of perturbations in the axion-gauge field system could thus turn out to moderately alter the quantitative results.

Finally, we have performed a fit of the model to an early time dataset only (Planck 18 and BAO). When comparing with the fit of $\Lambda$CDM to the same dataset, we find that the fluid description slightly improves the fit to the dataset, as was the case for~\cite{Poulin:2018cxd, Lin:2019qug}.


\section{Conclusions}
\label{sec:conclusions}

We explored a class of scenarios to alleviate the tension between early and late time measurements of the Hubble constant and its possible status within fundamental physics. The setup relies on ultralight scalars that can decay to massless fields, with a focus on the axion-dark-photon version. Soon after the start of scalar oscillations, the massless field/s can be resonantly enhanced due to tachyonic and parametric instabilities in its equation of motion. Therefore, energy can be efficiently transferred from the homogeneous scalar field to the massless field. This scenario represents a realization of a \emph{decaying ultralight scalar} (dULS) model.

By means of a linear analysis, we have provided evidence that sufficient resonant growth can occur in the first few scalar oscillations (in agreement with the lattice results obtained by~\cite{Kitajima:2017peg} for the case of the QCD axion). Because of the challenge of implementing numerically the resonant two-field system in existing Boltzmann codes, we have used an effective single fluid model to quantitatively assess the performance of this setup with respect to the Hubble tension. Our parameterization adds three extra parameters to the standard $\Lambda$CDM parameters: the relic abundance of the effective fluid and the values of the scale factor at which the two transitions from a dark energy to a matter-like fluid first, then to a radiation-like fluid, occur. These parameters correspond to the axion mass and decay constant and its coupling to the dark gauge fields. 

We have performed a fit to a combined early and late Universe dataset of a simplified version of the fluid model, obtained by fixing one of the three extra parameters and corresponding to an energy transfer from the axion field to the gauge fields which occurs during the first oscillation and at $z\sim 20,000$. We found $H_0=69.9_{-0.86}^{+0.84}~\text{km/s/Mpc}$ with $\Delta \chi^2\approx -9$ compared to the $\Lambda$CDM model, when the sound speed of the effective single fluid is taken to track the behavior of the equation of state parameter. For comparison, we find the $\Delta N_{\text{eff}}$ model to give very similar values of $H_0$ but  $\Delta \chi^2\approx -3$ compared to the $\Lambda$CDM model. Therefore, the fluid model performs better than $\Delta N_{\text{eff}}$, even when taking into account the larger number of new parameters.
Our results provide evidence that an axion field with a standard cosine potential and with $m\simeq 10^{-26}\text{eV}$ and $f\simeq 10^{17}~\text{GeV}$ can significantly alleviate the Hubble tension. However, even with the combined dataset considered in this work, the dULS model does not fully resolve the discrepancy between early and late time measurements of $H_0$. Furthermore, other data sets may worsen the fit, in particular those that constrain the $S_8$ parameter. In this respect, it would be interesting to investigate possibilities to suppress small-scale density fluctuations in our setup. These may require additional ingredients beyond the ones considered in this work, such as extra species with self interactions.

We also considered variations of our fluid parameterization, with a fixed value of the sound speed $c_s^2$. In particular, for $c_s^2=0.5$ we find $\Delta \chi^2\simeq -11$ with respect to $\Lambda$CDM, with a slightly larger value of $H_0$ compared to the case above.
Our modeling of the axion-gauge field system as a single fluid with varying equation of state parameter is certainly quite simple and may capture only some of the features of the field theory model. Nonetheless, the above results are promising and motivate future work towards a more realistic numerical implementation of the setup studied here.  

The dULS scenario can be compared to the EDE models studied in the literature.  From the point of view of fundamental physics, it offers an important advantage: it makes use of a standard axion potential which is quadratic around the minimum. This should be contrasted with EDE models, where a $V\propto(1-\cos(\phi/f))^n$, with $n\geq 2$ potential is assumed; this requires a miraculous conspiracy among coefficients of the harmonic expansion. Such a choice should take a big penalty in a fair $\Delta\chi^2$ analysis. Relatedly, such models currently have no known justification from a UV perspective. 

Furthermore, in its axionic version, the dULS model exploits the honest-to-goodness character of the axion: a pseudo-scalar with coupling to gauge fields $\phi F\tilde{F}/\M$. When considering a sound speed for fluid perturbations which tracks the equation of state parameter, we find that the Hubble tension is significantly alleviated when the energy transfer from the axion to the gauge field occurs during the first  oscillation. In this case, a dedicated numerical lattice analysis would be desirable to fully assess the relevance of backreaction effects on the axion field, which is left for future work. At the linear level, the regime of interest can be achieved by considering somewhat large axion-gauge field couplings $\M\lesssim f/80$. 

There is no known obstruction in making this choice of parameters from the point of view of effective field theory (since $\M$ controls the size of an operator that respects a symmetry, while $f$ controls the size of the axion field range). However, it is highly non-trivial to obtain a UV completion of this. Such UV completions have been argued to exist with additional ingredients, including extra fields (e.g. in the spirit of \cite{Kim:2004rp}, see also~\cite{Farina:2016tgd} for a clockwork model), but it is still an open question (e.g., see \cite{Agrawal:2018mkd}). 
Related issues apply to scalar-scalar models $\sim\phi\,\chi^2$, where the required parameters do not arise naturally from spontaneously broken quartic theories. Furthermore, direct perturbative decays require $\mathcal{O}(1)$ dimensionless couplings, which can introduce tuning of the ultralight scalar mass. 
In summary, our theoretical results, combined with possible problems from other data sets, are hints that the full resolution to the Hubble tension remains unsolved.

\section*{Acknowledgments}
We thank Thejs Brinckmann for correspondence on MontePython. We acknowledge use of Tufts HPC research cluster and thank its staff for help with the installation of MontePython. MPH is supported in part by National Science Foundation grants PHY-1720332, PHY-2013953.


%

\end{document}